\documentclass[doublecol]{epl2}

\title{The Equation of State for Solution of Semiflexible Polymer Chains}
\shorttitle{The Equation of State of Polymer Solution} 

\author{Yu.A. Budkov\inst{1} \and E.A. Nogovitsin \inst{2} \and A.L. Kolesnikov \inst{2}}
\shortauthor{Yu.A. Budkov \etal}

\institute{
  \inst{1} Institute of Solution Chemistry, Russian Academy of Sciences, 153045,
Ivanovo, Russia\\
  \inst{2} Ivanovo State University, Ermaka 39, 153025, Ivanovo, Russia}

\abstract{
We formulate a self-consistent procedure for calculation of thermodynamic and structural properties of polymer solutions based on the Gaussian equivalent representation method (GER) for functional integrals calculation beyond
the mean-field approximaton. We show that an equation of state, a potential of mean force of interaction monomer-monomer, and a persistent length should be defined self-consistently by solving of some system of coupled equation.}

\begin{document}

\maketitle

\section{Introduction}
As is well known, each polymer chain in solution interacts with large number of the same macromolecules that alows to replace the interaction
between chains by the interaction of latter with an effective self-consistent field. Approaches based on the idea on the self-consistent field appeared very fruitful to
description of the thermodynamic properties of the polymer solutions. By great efforts of Edwards, Helfand, and Tagamy functional methods get
into the polymer physics  \cite{Edwards,Helf}. The basic idea of Edwards that configuration space of long polymer chain looks like
a set of different paths of quantum particle in the framework of Feynman's formulation of quantum mechanics \cite{Fred}.   Thus, a
calculation of the polymer chain's partition function may be reduced to calculation of some path integral. Application of Hubbard-Stratonovich transformation allows to represent the
partition function of polymer solution in the form of functional integral on one or more fluctuating fields \cite{Fred}. The traditional methodology of functional integrals calculation in
polymer physics is mean-field (MF) theory which based on the saddle-point approximation \cite{Fred}. The MF theory gives a
quite good approximation of polymer solution equation of state in the high-concentrated regime only. At the MF level of approximation the monomer concentrtion fluctuations in the vicinity of saddle-points is usually
neglected. However, the latter assumption is incorrect in the case of semi-dilute polymer solution being characterized by the high fluctuations of monomer concentration \cite{Fred, Khohlov}. 
To carry out calculations for the dilute and semi-dilute regimes it is necessary to have the reliable methods describing thermodynamic properties of polymer solutions beyond the MF level of approximation \cite{Fred,Baeurle1}.

A successful attempt to go beyond MF-approximation was undertaken in works \cite{Muthu87, Muthu2001,Edwards_Muthu,Yethiraj, Donley1, Donley2, Brilliantov, Brilliantov2}. 
In these works a combination of variational method based on Gibbs-Bogolyubov inequality and random phase approximation (RPA) for calculation of thermodynamical properties of the flexible  polymer solution has been applied.

The Gaussian equivalent representation (GER) method is one of the approaches for calculations of the functional integrals beyond the MF level of
approximation and is a generalization of the variational method. The GER method has been developed and successfully applied in quantum
physics \cite{Efimov}. The zeroth approximation of GER is identical with result of a variational method. The corrections to
zeroth approximation can be calculated by the standard cumulant expansion method \cite{Kubo,PhysA}. As a rule, the second cumulant term no more than
five percent of the lowest approximation (the first correction is always equal to zero) \cite{Efimov, PhysA}. Besides, as distinct from the
variational, the GER method can be applied to the functional integrals with complex functionals. Presently, GER is actively used for a
description of thermodynamic and structural properties of polyelectrolyte solutions \cite{Baeurle2, Baeurle3, Baeurle4,2011,2012,Budkov,Budkov2}.

Semiflexible polymers are wide class of macromolecules that are include the important biopolymers. However, to the best of our knowledge, today there is only one attempt of a rigorous description of the structure properties of a nonideal semiflexible polymer chain in the solution, such as mean-square radius of gyration and effective persistent length \cite{MuthuSemiFlex}. Moreover, today there is no rigorous approaches to the theoretical calculations of the thermodynamic quantities of semiflexible polymer solutions from the first principles of statistical thermodynamics. In the work \cite{Budkov2} such self-consistent methodology, based on the GER method for the calculations of thermodynamic and structural properties of glycosaminoglycans aqueous solutions with additives of salt has been developed. However, the mathematical details of this methodology have been briefly discussed only. In this paper, we present the discussion of the mathematical details in a general form.

\section{Equation of state}
The grand canonical partition function for solution of semiflexible polymer chains has a following form
\begin{equation}
\Xi=\sum_{n=0}^{\infty}\frac{\xi^{n}}{n!}Q_{n},
\end{equation}
where
\begin{equation}
Q_{n}=\int\mathcal{D}\vec{r}_{1}..\int\mathcal{D}\vec{r}_{n} e^{-\beta\sum_{j=1}^{n}H_{0}[\vec{r}_{j}]-\beta H_{int}}
\end{equation}
is a canonical partition function for n semi-flexible polymer chains that are described by a following Hamiltonian \cite{MuthuSemiFlex,Netz}:
\begin{equation}
\beta H_{0}[\vec{r}]=\frac{3}{4l_{0}}\int_{0}^{L}ds(\dot{\vec{r}}^2(s)+l_{0}^2\ddot{\vec{r}}^2(s)).
\end{equation}

The total effective potential energy of pair interactions between monomers which should be renormalized by presence of the solvent has a form
\begin{equation}
\beta H_{int}=\frac{\beta}{2l_{0}^2}\sum_{i,j}\int_{0}^{L}\!\!\!\int_{0}^{L}ds_{1}ds_{2}U(\vec{r}_{i}(s_{1})-\vec{r}_{j}(s_{2})).
\end{equation}
It should be noted, here we consider the case of implicit solvent only.
The parameter $l_{0}$ is a persistent length of the ideal polymer chain (without interactions between monomers);
$\beta=\frac{1}{k_{B}T}$ is an inverse temperature; $L=Nl_{0}$ is the
length of polymer chain, and $N$ is effective number of segments
with length $l_{0}$. Using well known Hubbard-Stratonovich
transformation \cite{Fred}
\begin{equation}
\int\frac{\mathcal{D}\varphi}{C_{U}}e^{-\frac{1}{2}(\varphi U^{-1}\varphi)+i(b\varphi)}=e^{-\frac{1}{2}(b U b)},
\end{equation}
we represent the grand canonical partition function in the functional integral form as
\begin{equation}
\Xi=\int \frac{\mathcal{D}\varphi }{C_{U}}e^{-\frac{1}{2}(\varphi U^{-1}\varphi )+\xi\int \mathcal{D}\vec{r}e^{-\beta H_{0}[\vec{r}]+\frac{i\sqrt{\beta}}{l_{0}}\int_{0}^{L}ds\varphi (\vec{r}(s))}},
\end{equation}
where $C_{U}=\int\mathcal{D}\varphi e^{-\frac{1}{2}(\varphi U^{-1}\varphi)}$
is the normalization constant; inverse operator $U^{-1}$ can be determined by the following integral relation
$$\int d\vec{z} U(\vec{x}-\vec{z})U^{-1}(\vec{z}-\vec{y})=\delta^{(3)}(\vec{x}-\vec{y}),$$
where $\delta^{(3)}(\vec{x})$ is the three-dimensional Dirac delta function.
The following short-hand notation have been entered:
$$(\varphi U^{-1}\varphi )=\int d\vec{x}\int d\vec{y}\varphi (\vec{x})U^{-1}(\vec{x}-\vec{y})\varphi (\vec{y}).$$
Let us apply GER method to calculation of the functional integral (5). Performing the displacement of the functional variable
\begin{equation}
\varphi \rightarrow \varphi +\frac{i\varphi _{0}}{\sqrt{\beta}},
\end{equation}
going to a new Gaussian measure with a new propagator $D(\vec{x}-\vec{y})$,  one can obtain the following representation for the grand partition function 
\begin{equation}
\Xi=\frac{C_{D}}{C_{U}}\int\frac{\mathcal{D}\varphi }{C_{D}}e^{-\frac{1}{2}(\varphi D^{-1}\varphi )+W_{D}[\varphi ]},
\end{equation}
where $C_{D}=\int\mathcal{D}\varphi e^{-\frac{1}{2}(\varphi D^{-1}\varphi)}$ is a normalization constant for the new Gaussian measure;
$$W_{D}[\varphi]=-\frac{1}{2}:(\varphi [U^{-1}-D^{-1}]\varphi ):_{D}-\frac{1}{2}tr(D[U^{-1}-D^{-1}])$$
$$-\frac{i}{\sqrt{\beta}}(\varphi _{0}U^{-1}\varphi )+\frac{1}{2\beta}(\varphi _{0}U^{-1}\varphi _{0})$$
\begin{equation}
+\bar{\xi} e^{-N\varphi _{0}}\int d\sigma[\vec{r}]:e^{\frac{i\sqrt{\beta}}{2l_{0}}\int_{0}^{L}ds\varphi (\vec{r}(s))}:_{D},
\end{equation}
$$\int d\sigma[\vec{r}][*]=\frac{\int\mathcal{D}\vec{r}e^{-\beta H_{0}[\vec{r}]-\frac{\beta}{2l_{0}^2}\int_{0}^{L}\!\!\!\int_{0}^{L}ds_{1}ds_{2}D(\vec{r}(s_{1})-\vec{r}(s_{2}))}[*]}{\int\mathcal{D}\vec{r}e^{-\beta H_{0}[\vec{r}]-\frac{\beta}{2l_{0}^2}\int_{0}^{L}\!\!\!\int_{0}^{L}ds_{1}ds_{2}D(\vec{r}(s_{1})-\vec{r}(s_{2}))}},$$
is a "functional of interaction"
and
$$\bar{\xi}=\xi\int\mathcal{D}\vec{r}e^{-\beta H_{0}[\vec{r}]-\frac{\beta}{2l_{0}^2}\int_{0}^{L}\!\!\!\int_{0}^{L}ds_{1}ds_{2}D(\vec{r}(s_{1})-\vec{r}(s_{2}))}$$
is renormalized activity. We have also introduced the concept of the normal product according to the new Gaussian measure with the new propagator $D(\vec{x}-\vec{y})$. 
Accounting of normal product leads to a summation of a so-called tadpole diagrams \cite{Efimov}.

The following short-hand notations have been introduced:
$$:e^{\frac{i\sqrt{\beta}}{l_{0}}\int_{0}^{L}ds\varphi (\vec{r}(s))}:_{D}=$$
\begin{equation}
=e^{\frac{i\sqrt{\beta}}{l_{0}}\int_{0}^{L}ds\varphi (\vec{r}(s))}e^{\frac{\beta}{2l_{0}^2}\int_{0}^{L}\!\!\!\int_{0}^{L}ds_{1}ds_{2}D(\vec{r}(s_{1})-\vec{r}(s_{2}))},
\end{equation}
\begin{equation}
:\varphi (\vec{x})\varphi (\vec{y}):_{D}=\varphi (\vec{x})\varphi (\vec{y})-D(\vec{x}-\vec{y}),
\end{equation}
\begin{equation}
\int \frac{\mathcal{D}\varphi }{C_{D}}e^{-\frac{1}{2}(\varphi D^{-1}\varphi )}:e^{\frac{i\sqrt{\beta}}{l_{0}}\int_{0}^{L}ds\varphi (\vec{r}(s))}:_{D}=1,
\end{equation}
$$tr\left(D[U^{-1}-D^{-1}]\right)=$$
\begin{equation}
=\int d\vec{x}\int d\vec{y}D(\vec{x}-\vec{y})\left[U^{-1}(\vec{y}-\vec{x})-D^{-1}(\vec{y}-\vec{x})\right].
\end{equation}
The basic idea of the GER method is that the main contribution to the functional integral is $"$concentrated$"$ in the new Gaussian measure \cite{Efimov,PhysA}.
This means that the linear and quadratic terms over the integration variable $\varphi(\vec{x})$ should be absent in the integrand.
Thus, we obtain two conditions
\begin{equation}
-\frac{i}{\sqrt{\beta}}(\varphi _{0}U^{-1}\varphi )+\frac{i\sqrt{\beta}}{l_{0}}\bar{\xi}e^{-N\varphi _{0}}\int_{0}^{L}ds\left<\varphi (\vec{r}(s))\right>=0,
\end{equation}
$$-\frac{1}{2}(\varphi [U^{-1}-D^{-1}]\varphi )$$
\begin{equation}
-\frac{\bar{\xi}\beta e^{-N\varphi _{0}}}{2 l_{0}^2}\int_{0}^{L}\!\!\!\int_{0}^{L}ds_{1}ds_{2}\left<\varphi (\vec{r}(s_{1}))\varphi (\vec{r}(s_{2}))\right>=0,
\end{equation}
that are valid at any field configurations $\varphi (\vec{x})$.
Further, excluding the variables $\varphi$, we arrive at two self-consistent equations for calculation of the new propagator $D(\vec{x})$ and parameter of displacement $\varphi_{0}$
\begin{equation}
\label{eq:phi0}
\varphi _{0}=\beta \bar{\xi}N\tilde{U}(0)e^{-N\varphi _{0}},
\end{equation}
\begin{equation}
\tilde{D}(\vec{k})=\frac{\tilde{U}(\vec{k})}{1+\varphi _{0}G(\vec{k})\frac{\tilde{U}(\vec{k})}{\tilde{U}(0)}},
\end{equation}
where $\tilde{D}(\vec{k})$ and $\tilde{U}(\vec{k})$ are Fourier-images of functions $D(\vec{x})$ and $U(\vec{x})$,
respectively.
The following short-hand notation has been also introduced:
\begin{equation}
\left<(\cdot )\right>=\int d\sigma[r](\cdot ),
\end{equation}
and
\begin{equation}
G(\vec{k})=\frac{1}{l_{0}L}\int_{0}^{L}\!\!\!\int_{0}^{L}ds_{1}ds_{2}\left<e^{i\vec{k}(\vec{r}(s_{1})-\vec{r}(s_{2}))}\right>
\end{equation}
is a static structure factor of the polymer chain in the solution \cite{Flory_book,Khohlov}.
Thus, the grand partition function can be rewritten in the following Gaussian equivalent form
\begin{equation}
\label{eq:Xi}
\Xi =e^{\beta PV}=e^{\beta P_{0}V}\int \frac{\mathcal{D}\varphi }{C_{D}}e^{-\frac{1}{2}(\varphi D^{-1}\varphi )+W_{2}[\varphi ]}.
\end{equation}
It should be noted, that the linear and quadratic terms over the integration variable $\varphi (x)$ are absent in a functional $W_{2}[\varphi ]$.
The expression for an osmotic pressure of the polymer solution in zeroth approximation has a following form
\begin{equation}
\label{eq:P0}
\frac{P_{0}}{k_{B}T}=\frac{2\varphi _{0}+N\varphi _{0}^2}{2N\beta \tilde{U}(0)}+\frac{\varphi _{0}^2}{12\pi^2}\int_{0}^{\infty}\frac{dk k^3 u(k)u^{\prime}(k)}{(1+\varphi _{0}u(k))^2}.
\end{equation}
The average polymer chains density can be calculated by the standard thermodynamic identity:
\begin{equation}
\label{eq:rho_ch}
\left<\rho \right>_{GC}=\frac{\bar{\xi}}{V}\left(\frac{\partial\log{\Xi}}{\partial{\bar{\xi}}}\right)_{T}.
\end{equation}
Using the relation (\ref{eq:phi0})  and relations (\ref{eq:Xi}-\ref{eq:rho_ch}), we obtain in zeroth approximation of GER
\begin{equation}
\label{eq:rho}
\rho _{m}=\frac{\varphi _{0}}{\beta\tilde{U}(0)}-\frac{N\varphi _{0}^2}{2\pi(1+N\varphi _{0})}\int_{0}^{\infty}\frac{dk k^2u^2(k)}{(1+\varphi _{0}u(k))^2},
\end{equation}
where $u(k)=G(\vec{k})\frac{\tilde{U}(\vec{k})}{\tilde{U}(0)}$, $\rho_{m}=\left<\rho \right>_{GC}N$. 
We would like to stress, that in contrast to the pure MF approximation, within present approach there is an additional second term in the right hand side of (\ref{eq:rho}) which is related to the correlations 
between monomers beyond the MF approximation. As it follows from equation (\ref{eq:rho}) the MF approximation overestimates a value of the monomer concentration.

The equations (\ref{eq:P0}) and (\ref{eq:rho}) define the equation of state for polymer solution in parametrical form.

\section{Static structure factor in GER}
To calculate the osmotic pressure as a function of the monomer concentration and temperature it is necessary to have 
an expression for the static structure factor
\begin{equation}
\label{eq:G}
G(\vec{k})=\frac{1}{l_{0}L}\int_{0}^{L}\!\!\!\int_{0}^{L}ds_{1}ds_{2}\left<e^{i\vec{k}(\vec{r}(s_{1})-\vec{r}(s_{2}))}\right>,
\end{equation}
where
$$\left<e^{i\vec{k}(\vec{r}(s_{1})-\vec{r}(s_{2}))}\right>=\int d\sigma[\vec{r}]e^{i\vec{k}(\vec{r}(s_{1})-\vec{r}(s_{2}))}=$$
\begin{equation}
=\int\frac{\mathcal{D}\vec{r}}{Z}e^{-\beta H_{0}[\vec{r}]-\frac{\beta}{2l_{0}^2}\int_{0}^{L}\!\!\!\!\int_{0}^{L}ds ds'D(\vec{r}(s)-\vec{r}(s'))}e^{i\vec{k}(\vec{r}(s_{1})-\vec{r}(s_{2}))},
\end{equation}
\begin{equation}
Z=\int\mathcal{D}\vec{r}e^{-\beta H_{0}[\vec{r}]-\frac{\beta}{2l_{0}^2}\int_{0}^{L}\!\!\!\int_{0}^{L}ds_{1}ds_{2}D(\vec{r}(s_{1})-\vec{r}(s_{2}))}
\end{equation}
is a partition function of a single polymer chain in the solution.
We would like to stress that the function $D(\vec{r})$ has a sense of the potential of mean force of interaction between monomers \cite{2011,2012}.
The partition function $Z$ can be written as a functional integral over the Fourier-components of random functions $\vec{r}(s)=\int_{-\infty}^{\infty}\frac{dq}{2\pi}\vec{r}_{q}e^{iqs}$:
$$Z=\int\prod_{q\neq 0}\frac{d\vec{r}_{q}d\vec{r}_{-q}}{(2\pi g_{0}(q))^{3/2}}e^{-\frac{1}{2}\int_{-\infty}^{\infty}\frac{dq}{2\pi}\frac{\vec{r}_{q}\vec{r}_{-q}}{g_{0}(q)}}\times$$
\begin{equation}
\times e^{-\frac{\beta}{2l_{0}^2}\int_{0}^{L}\!\!\!\int_{0}^{L}ds_{1}ds_{2}\int\frac{d\vec{k}}{(2\pi)^3}\tilde{D}(\vec{k})e^{i\vec{k}\int_{-\infty}^{\infty}\frac{dq}{2\pi}\vec{r}_{q}(e^{iqs_{1}}-e^{iqs_{2}})}},
\end{equation}
where
\begin{equation}
g_{0}(q)=\frac{2l_{0}}{3q^2(1+l_{0}^2q^2)}.
\end{equation}
Let us represent the partition function $Z$ in the GER form.
In this case there is no need to do the shift of integration variable because functional of interaction is simmetric one \cite{Efimov}.
Let us go over to a new Gaussian measure with the new propagator $g(q)$. Thus
\begin{equation}
Z=e^{W_{0}}\int\prod_{q\neq 0}\frac{d\vec{r}_{q}d\vec{r}_{-q}}{(2\pi g(q))^{3/2}}e^{-\frac{1}{2}\int_{-\infty}^{\infty}\frac{dq}{2\pi}\frac{\vec{r}_{q}\vec{r}_{-q}}{g(q)}+W_{I}[\vec{r}]},
\end{equation}
where
\begin{equation}
W_{0}=\frac{3L}{2}\int_{-\infty}^{\infty}\frac{dq}{2\pi}\left[\log{\frac{g(p)}{g_{0}(q)}}-\frac{g(q)}{g_{0}(q)}+1\right],
\end{equation}
and
$$W_{I}[\vec{r}]=-\frac{1}{2}\int_{-\infty}^{\infty}\frac{dq}{2\pi}:\vec{r}_{q}\vec{r}_{-q}:_{g}\left(\frac{1}{g_{0}(p)}-\frac{1}{g(p)}\right)$$
$$-\frac{\beta}{2l_{0}^2}\int_{0}^{L}\!\!\!\!\!\int_{0}^{L}ds_{1}ds_{2}\int\frac{d\vec{k}}{(2\pi )^3}\tilde{D}(\vec{k})$$
\begin{equation}
\times e^{-k^2\int_{-\infty}^{\infty}\frac{dq}{\pi}g(q)\sin^{2}\frac{q(s_{1}-s_{2})}{2}} :e^{i\vec{k}\int_{-\infty}^{\infty}\frac{dq}{\pi}\vec{r}_{q}(e^{iqs_{1}}-e^{iqs_{2}})}:_{g}.
\end{equation}

The following notations have been introduced:

$$:e^{i\vec{k}(\vec{r}(s_{1})-\vec{r}(s_{2}))}:_{g}=e^{i\vec{k}(\vec{r}(s_{1})-\vec{r}(s_{2}))}$$
\begin{equation}
\times e^{-k^2\int_{-\infty}^{\infty}\frac{dq}{\pi}g(q)\sin^{2}\frac{q(s_{1}-s_{2})}{2}},
\end{equation}

\begin{equation}
:\vec{r}_{q}\vec{r}_{q'}:_{g}=\vec{r}_{q}\vec{r}_{q'}-6\pi g(q)\delta(q+q').
\end{equation}

The basic contribution to value of functional integral (28) is
concentrated in a new Gaussian measure with the propagator $g(q)$.
It means that quadratic terms over variables $\vec{r}_{q}$ in
$W_{I}[\vec{r}]$ are absent. It is corresponding to a following
condition:

\begin{equation}
\frac{1}{g_{0}(q)}-\frac{1}{g(q)}=\Sigma (q|g),
\end{equation}

which we can represent in the following form:

\begin{equation}
\label{eq:g(q)}
g(q)=\frac{g_{0}(q)}{1-\Sigma (q|g)g_{0}(q)},
\end{equation}

where
$$\Sigma (q|g)=\frac{2\beta}{3l_{0}^2L}\int_{0}^{L}\!\!\!\int_{0}^{L}ds_{1}ds_{2}\sin^{2}\frac{q(s_{1}-s_{2})}{2}$$
\begin{equation}
\times \int\frac{d\vec{k}}{(2\pi)^3}k^2\tilde{D}(\vec{k})e^{-k^2\int_{-\infty}^{\infty}\frac{dq}{\pi}g(q)\sin^{2}\frac{q(s_{1}-s_{2})}{2}},
\end{equation}

The nonlinear integral equation (\ref{eq:g(q)}) can be solved with respect to $g(q)$, for example, by iterative method.

At the large scales, i.e. at $q\rightarrow 0$, we have the following approximated expression for propagator:
\begin{equation}
g(q)\simeq \frac{2l_{1}}{3q^2(1+l_{1}^2q^2)},
\end{equation}
where $l_{1}$ is a renormalized persistent length of the polymer chain.
The equation for calculation of the renormalized persistent length $l_{1}$ has a form:
$$\frac{1}{l_{0}}-\frac{1}{l_{1}}=\frac{\beta}{9Ll_{0}^2}\int_{0}^{L}\!\!\!\int_{0}^{L}ds_{1}ds_{2}(s_{1}-s_{2})^2$$
\begin{equation}
\label{eq:l1}
\times \int\frac{d\vec{k}}{(2\pi )^3}k^2\tilde{D}(\vec{k})e^{-\frac{2k^2l_{1}^2}{3}\left(\frac{|s_{1}-s_{2}|}{l_{1}}+e^{-\frac{|s_{1}-s_{2}|}{l_{1}}}-1\right)}.
\end{equation}

Thus we have the following approximated expression for structure factor of polymer chain:
$$G(\vec{k})\simeq \frac{1}{l_{0}L}\int_{0}^{L}\!\!\!\int_{0}^{L}ds_{1}ds_{2}\frac{\int\mathcal{D}\vec{r}e^{-\beta H_{1}[\vec{r}]}e^{i\vec{k}(\vec{r}(s_{1})-\vec{r}(s_{2}))}}{\int\mathcal{D}\vec{r}e^{-\beta H_{1}[\vec{r}]}}=$$
\begin{equation}
\label{eq:G}
=\frac{1}{l_{0}L}\int_{0}^{L}\!\!\!\int_{0}^{L}ds_{1}ds_{2}e^{-\frac{2k^2l_{1}^2}{3}\left(\frac{|s_{1}-s_{2}|}{l_{1}}+e^{-\frac{|s_{1}-s_{2}|}{l_{1}}}-1\right)},
\end{equation}
where
\begin{equation}
\beta H_{1}[\vec{r}]=\frac{3}{4l_{1}}\int_{0}^{L}ds(\dot{\vec{r}}^2(s)+l_{1}^2\ddot{\vec{r}}^2(s)).
\end{equation}

At $k\rightarrow 0$
\begin{equation}
G(\vec{k})\simeq  N\left(1-\frac{1}{3}R_{g}^2k^2\right),
\end{equation}
where $R_{g}$ is a radius of gyration of the polymer chain \cite{Khohlov}. After some calculations we obtain a well known 
expression for the radius of gyration of semiflexible polymer chain
\begin{equation}
R_{g}^2=\frac{Ll_{1}}{3}-l_{1}^2+\frac{2l_{1}^3}{L}-\frac{2l_{1}^4}{L^2}\left(1-e^{-\frac{L}{l_{1}}}\right)
\end{equation}
with the persistent length $l_{1}$ which implicitly depends on parameters of interaction, temperature and monomer concentration.

\section{Conclusion}
In summary, we have obtained the closed system of coupled equations ((\ref{eq:P0}),
(\ref{eq:phi0}), (\ref{eq:l1}), (\ref{eq:G})), which define the equation of state for the system of semiflexible polymer chains via the pair monomers
interactions (4). In this short article we does not specify these interactions. The choice of the form of the potential of interaction is a specific target 
\cite{2012,Budkov,Budkov2}.

The important task which remains still unsolved is a solution of nonlinear integral equation (\ref{eq:g(q)}). The solution of this equation will allow to obtain the information 
on a structure of the semiflexible polymer chain in the solution for a wide range of a length scale. The numerical calculation of this integral equation is a subject of the future research.

\end{document}